\documentclass[12pt]{iopart}
\usepackage{braket}
\usepackage[unicode=true,pdfusetitle,
 bookmarks=true,bookmarksnumbered=false,bookmarksopen=false,
 breaklinks=false,pdfborder={0 0 0},pdfborderstyle={},backref=false,colorlinks=true]
 {hyperref}
 \hypersetup{
 citecolor=blue,
 urlcolor=blue,
 linkcolor=blue}

\expandafter\let\csname equation*\endcsname\relax
\expandafter\let\csname endequation*\endcsname\relax
\usepackage{amsmath}
\usepackage{latexsym,verbatim}
\usepackage{amssymb} 
\usepackage{amsfonts}
\usepackage{colortbl}
\usepackage{array}
\usepackage{stackrel}
\usepackage{bm}
\usepackage{nicefrac}
\usepackage{rotating}
\usepackage{float}
\usepackage[final]{pdfpages}
\usepackage{fancyhdr}

\newcommand{\new}[1]{{\color{black}{#1}}}

\begin{document}

\title[Laplacian Renormalization Group: a practical overview]{Laplacian Renormalization Group: An introduction to heterogeneous coarse-graining}


\author{Guido Caldarelli\textsuperscript{1,2,3,4}, Andrea Gabrielli\textsuperscript{3,5,6}, Tommaso Gili\textsuperscript{7}, Pablo Villegas\textsuperscript{5,8}}

\address{$^1$ DMSN, Ca'Foscari University of Venice, Via Torino 155, 30172 - Venice, Italy}
\address{$^2$ European Centre for Living Technology (ECLT), Dorsoduro 3911, 30123 - Venice, Italy}
\address{$^3$ Institute for Complex Systems (ISC), CNR, UoS Sapienza, Piazzale Aldo Moro 2, 00185 - Rome, Italy}
\address{$^4$ London Institute for Mathematical Sciences (LIMS), W1K2XF London, United Kingdom}
\address{$^5$ `Enrico Fermi' Research Center (CREF), Via Panisperna 89A, 00184 - Rome, Italy}
\address{$^6$ Dipartimento di Ingegneria Civile, Informatica e delle Tcnologie Aeronautiche, Universit\`a degli Studi `Roma Tre', Via Vito Volterra 62, 00146 - Rome, Italy}
\address{$^7$ Networks Unit, IMT Scuola Alti Studi Lucca, Piazza San Francesco 15, 55100- Lucca, Italy}
\address{$^8$ Instituto Carlos I de F\'isica Te\'orica y Computacional, Universidad de Granada, Granada, Spain}

\ead{guido.caldarelli@unive.it, andrea.gabrielli@cref.it, tommaso.gili@imtlucca.it, pablo.villegas@cref.it}

\vspace{10pt}

\begin{abstract}

The renormalization group (RG) constitutes a fundamental framework in modern theoretical physics. It allows the study of many systems showing states with large-scale correlations and their classification in a relatively small set of universality classes. RG is the most powerful tool for investigating organizational scales within dynamic systems. However, the application of RG techniques to complex networks has presented significant challenges, primarily due to the intricate interplay of correlations on multiple scales.

Existing approaches have relied on hypotheses involving hidden geometries and based on embedding complex networks into hidden metric spaces. Here, we present a practical overview of the recently introduced Laplacian Renormalization Group for heterogeneous networks. First, we present a brief overview that justifies the use of the Laplacian as a natural extension for well-known field theories to analyze spatial disorder. We then draw an analogy to traditional real-space renormalization group procedures, explaining how the LRG generalizes the concept of "Kadanoff supernodes" as block nodes that span multiple scales. These supernodes help mitigate the effects of cross-scale correlations due to small-world properties. Additionally, we rigorously define the LRG procedure in momentum space in the spirit of Wilson RG. Finally, we show different analyses for the evolution of network properties along the LRG flow \new{following structural changes when the network is properly reduced}.
\end{abstract}

\tableofcontents{}
\newpage

\section{Introduction} 

\indent In 1911, Paul and Tatiana Ehrenfest introduced the notion of coarse-graining in many-body systems \cite{ehrenfets1911begriffliche}. The idea was to introduce an operation to transform a probability density in phase space into a “coarse-grained” density, a piece-wise constant function resulting from density averaging in small but finite cells. Many decades later, the idea of coarse-graining had a dramatic impact everywhere in statistical physics (both equilibrium and non-equilibrium). In particular, it was the central idea of the Kadanoff transformation. In 1966,  Leo P. Kadanoff applied the concept of coarse-graining for statistical mechanical systems close to a second-order phase transition point to introduce the concepts of the scaling theory and ``block-spin" renormalization group for this class of systems \cite{kadanoff1966scaling,kadanoff1976}. This innovative approach, known as the ``blocking idea", provides a method for characterizing the components of a theory at arbitrarily large resolution scales by considering them as collective combinations of components at smaller distances and can be considered as a precursive conceptualization of the Renormalization Group (RG) theory introduced in a field theory formalism by Kenneth Wilson \cite{wilson1974renormalization} (see also \cite{di1969microscopic}) both as a microscopic foundation and a natural approach to scaling theories in critical phenomena, and a tool for calculating the system properties around the critical point. In its original formulation proposed by Gell-Mann and Low \cite{gell1954quantum}, this apparatus was a simple mathematical trick to deal with the appearance of infinities in quantum field perturbation theories. Thanks mainly to Kadanoff's and Wilson's work, it became a solid theory with a deep physical meaning rapidly adopted to describe equilibrium critical phenomena \cite{zinn2021quantum}. 
%

In the last twenty years, the RG field of applicability has been 
extended to out-of-equilibrium systems defined in homogeneous spaces, i.e., regular lattices, such as contact processes (e.g., epidemic models), dynamical and directed percolation, opinion dynamics (e.g., voter models) and models for synchronization of oscillators (e.g., Kuramoto model of non-linear coupled oscillators) \cite{Hinrichsen00, Dornic2005}. 
This large class of systems has proven useful to model many real-world phenomena, where, however, the pattern of interactions between system constituents can rarely be represented by a regular lattice. Instead, they more properly form an irregular network characterized by heterogeneous topology and geometry. A modern version of coarse-graining and renormalization 
deals with topology, particularly the contacts (whatever they could represent according to the different many-body interaction scenarios considered) that one element in a system may have.
The branch of mathematics able to describe such systems 
is graph theory \cite{bollobas1998modern}, where vertices and edges represent the elements and their connections, respectively.
The declination of graph theory in the diverse schemes of many-body interactions gave place to the complex network theory 
\cite{caldarelli2007scale,cimini2019statistical} that now represents an almost universal way to model a variety of processes from fake news diffusion \cite{lazer2018science} in the society \cite{caldarelli2018physics} to financial markets \cite{bardoscia2021physics}, urban development \cite{caldarelli2023role}, software \cite{Debian}, medicine \cite{barabasi2011network}, and ecology \cite{villegas2021emergent}.
In network representation of real systems, nodes represent the constituent elements connected by links representing their interactions. In this way, a graph provides a topological network structure characterized by complex connectivity patterns with coupled short- and long-range connections. These structures sustain dynamical processes, defined by appropriate node variables, typically represented by the models mentioned above, defining the time-varying states of nodes. Also, in this {\em heterogeneous} case, at certain values of the model parameters, transitions from noisy decoupled states (e.g., no macroscopic epidemic spread) to collective critical ones (e.g., macroscopic epidemic spread) are observed. To study these transitions, one would like to generalize the RG approach to the case of irregular networks.
However, core assumptions that underpin the successes of the ordinary RG approach, such as locality and spatial homogeneity of interactions, no longer apply to systems defined on complex networks.
Instead, the extremely heterogeneous, small-world, and multiscale architecture of interactions cannot be treated as a mere topological perturbation of simple homogeneous cases but demands an entirely new multiscale theory that combines {\em ab initio} the complex topological structure of the networked space with the “physical” processes running on it. Transitions arise from the complex interplay between the irregular space topology and the parameters of the dynamical process. A general and formal theory for the complex interplay between the coupled multiscale structures of real networks and the “physical” interactions between the elements of the system determining the formation of critical collective states or “phases” is still an open problem, despite its central importance in modern science and society \cite{NPEd2023}. Indeed the lack of topological homogeneity and the small world features of the embedding space make the definition of the coarse-graining procedure at the base of the RG approach quite problematic and arbitrary: it is unclear how to define equivalent neighborhoods for the different nodes at an arbitrary scale.
This problem can be rephrased as the difficulty of defining appropriate spaces of representation of networks in which the concepts of locality and proximity can be recovered to define equivalent neighborhoods of arbitrary "diameter"  of network nodes, which are the main ingredient in implementing the coarse-graining procedure and the consequent RG scheme. 

\new{Numerous efforts have been undertaken to elucidate the transition from microscopic to macroscopic properties in complex networks \cite{newman1999renormalization,song2005self,radicchi2009renormalization, rozenfeld2010small,aygun2011spectral}. Here is a concise overview of the key studies. Newman et al. \cite{newman1999renormalization} studied the small-world network model of Watts and Strogatz using an asymptotically exact real-space renormalization group method. They found that in all dimensions $d$ the model undergoes a continuous phase transition as the density $p$ of shortcuts tends to zero and that the characteristic length $\xi$ diverges according to $\xi \sim \rho^{-\tau}$ with $\tau = 1/d$.
Song et al. \cite{song2005self} proposed transforming a network using a box-covering technique, in which a box includes nodes such that the distance between each pair of nodes within a box is smaller than a threshold $l_{B}$. After tiling the network, the nodes of each box and their mutual links are replaced by a supernode: supernodes are connected if there is at least one link between the nodes of their corresponding boxes in the original network. This defines a renormalization transformation $R_{l_{B}}$. For some real networks, such as the WWW, social, metabolic and protein-protein interaction networks, a few iterated applications of this procedure leave their degree distribution invariant, which led to the claim that they are self-similar. 
Radicchi et al. \cite{radicchi2009renormalization} proposed iterated applications of  $R_{l_{B}}$ to generate renormalization flows in the space of all possible graphs. They showed that renormalization flows in graphs, as defined by the box-covering method, are similar to the renormalization of spin systems, leveraging on the analysis of classic renormalization for percolation and the Ising model on the lattice. 
Accordingly, Rozenfeld et al. \cite{rozenfeld2010small} showed that the RG flow readily identifies a small-world/fractal transition by finding a trivial stable fixed point of a complete graph, a non-trivial point of a pure fractal topology that is stable or unstable according to the number of long-range links in the network, and another stable point of a fractal with short-cuts that exists exactly at the small-world/fractal transition. Finally, Ayg\"un et al. \cite{aygun2011spectral} explored the possibilities offered by the eigenvectors and eigenvalues of the graph Laplacian to develop a field theoretic renormalization group approach to order-disorder phenomena on complex networks. In brief, they expanded order parameter fluctuations in eigenvectors of the graph Laplacian, wrote down the equivalent of a Ginzburg-Landau Hamiltonian, and then performed partial summations over the partition function, to eliminate the high-eigenvalue components. They showed that the proliferation of higher-order terms in the renormalized Hamiltonian was controlled by going over to a quenched average over different realizations of the stochastic network using a replica approach.} Recently, two more founded approaches have been proposed: (i) Geometric Renormalization \cite{boguna2021network, garcia2018multiscale} based on embedding networks into underlying hidden metric spaces, and (ii) stochastic ensemble renormalization \cite{garlaschelli2023}.


Here, we present a comprehensive summary of the recently proposed dynamical and geometry-free theoretical approach to define an RG for complex networks \cite{villegas2023laplacian}, based on a fundamental dynamical process evolving on the graph structure: the Laplacian diffusion of information among nodes \cite{dedomenico2016spectral,masuda2017random} governed by the Laplacian evolution operator of the graph.
To begin with, we establish an intuitive real-space variant of the Renormalization Group (RG), drawing inspiration from the Migdal-Kadanoff RG approach \cite{migdal1976phase, kadanoff1976}. This framework introduces a recursive procedure for coarsening network nodes ({\em i.e., decimation}) while preserving their diffusion characteristics at progressively larger spatiotemporal scales. Following the principles of real-space RG techniques \cite{kadanoff1966scaling}, we introduce the concept of Kadanoff supernodes, guided by the inherent resolution scales of the system. This approach effectively addresses issues related to small-world networks and efficiently resolves decimation problems when constructing downscaled replicas.

Subsequently, we move towards a more rigorous formulation of the diffusion-driven RG, akin to the $k$-space RG, in the spirit of Wilson's approach in statistical field theory. This permits us to establish a full proposal of a novel Laplacian RG (LRG) theoretical framework, wherein fast diffusion modes are systematically integrated out from the Laplacian operator, analogous to the conjugate Fourier space. In this way, we can afford the problem of scale transformation and renormalization \new{opening the door to future applications in particular} dynamical processes for which a field theory-like representation is possible (e.g., contact process). Finally, we present here, for the first time, different analyses of the evolution of different network metrics on the LRG flow and show different coarse-grainings of weighted networks that can be of utmost importance for different biological applications.

\new{The structure of the paper is as follows. Section 2 briefly reviews the fundamental aspects of classical field theories, initially describing some general concepts and connecting them with the need for a complete RG formulation of systems with quenched disorder. The familiar reader can safely skip this section and move to Section 2.3. In particular, we present previous results legitimating the connection of the Gaussian model with the Laplacian of any network structure. In section 3, we present the main points of the Laplacian Renormalization Group (LRG), emphasizing the importance of performing coarse-graining in heterogeneous systems. Moreover, we present different examples of how different structural properties evolve under coarse-grain transformations. Finally, we present a broader discussion of our results.}

\section{Methods}
\subsection{Zooming out scales in field theories}

Let us start with an overarching perspective of the issue. We refer, however, to \cite{kardar2007statistical,amit2005field,zinnjustin2007phase,binney1992theory} for a detailed description of the framework. We first focus on physical systems embedded in homogeneous spaces (i.e., $d-$dimensional Euclidean space or regular lattice) with local and homogeneous interactions whose thermal fluctuations near the transition temperature can be described by a scalar field \cite{Schmittmann1995, Marro}. In particular, we begin defining the bare order parameter $\phi=\phi(\mathbf{r},t)$ and its average over time corresponding to $\langle \phi(\mathbf{r},t) \rangle$. Specifically, we pivot towards \new{the so-called} Model B in Hohenberg-Halperin \cite{HH}, which describes the Ising model for ferromagnetism. Note that, in this specific case, the generalized Landau-Ginzburg Hamiltonian reads,
\begin{equation}
    \mathcal{H}\left[\phi\right]=\int d^{d}\mathbf{x}\left\{ \frac{1}{2}\left(\vec{\nabla}\phi\right)^{2}+\frac{1}{2!}\mu^{2}\phi^{2}+\frac{\lambda}{4!}\phi^{4}+\frac{\kappa}{6!}\phi^{6}+\ldots\right\},
    \label{Ising}
\end{equation}
and the associated Langevin equation as,
\begin{equation}
\dot{\phi}=\nabla_{i}\left\{\gamma\nabla_{i}\frac{\delta\mathcal{H}}{\delta\phi}+\xi_{i}\right\},
\end{equation}
where $\mu^2=T-T_c$ represents the distance to the critical temperature of the system, $\left(\vec{\nabla}\phi\right)^{2}$ describes the diffusive coupling of the scalar field
with its nearest neighbors, and $\xi_i$ is Gaussian white noise with zero mean, and ``delta-correlated": $\langle\xi(\mathbf{x},t)\xi(\mathbf{x^\prime},t^\prime)\rangle\propto\delta(\mathbf{x}-\mathbf{x^\prime})\delta(t-t^\prime)$. It is important to underline that the first two quadratic terms in the parenthesis in Eq.~\eqref{Ising} define the so-called {\em Gaussian approximation} for the statistical field theory, which is exactly solvable through Gaussian functional integrals \new{(we will discuss below its extension to heterogeneous structures)}. The minimum (Euclidean) spatial dimension $d_c$ above which the Gaussian approximation gives the same solution of the complete model is called the `upper critical dimension' and plays a fundamental role in the RG approach.

Notably, it is possible to generalize this field-theoretical description for different out-of-equilibrium models. The contact process (CP) is an instance of this, where, in the spirit of the time-dependent Landau-Ginzburg method for critical dynamics \cite{HH, Marro}, one can formulate the general Langevin equation,
\begin{equation}
\dot{\phi}=a\phi-b\phi^2-c\phi^3+\ldots+D\nabla^2\phi+\sqrt{\phi}\xi(\mathbf{x},t)
\label{CP}
\end{equation}
where $\phi(\mathbf{x},t)$ is the local order parameter now representing a density of active states, and $\xi$ is a Gaussian white noise with zero mean and delta-correlated. If the system is assumed statistically symmetric under reflection, the even power terms in Eq.~\eqref{CP} are not present.
 
In principle, one can consider the expansion up to an arbitrary coupling of the form $g_n\phi^n$ either in Eq.\eqref{Ising} and Eq.\eqref{CP} (under the constraint of respecting the physical symmetries defining the underlying problem). The number of significant or \emph{relevant} couplings, i.e. that do not vanish under the RG flow defined below depending on the spatial dimension of the system \cite{amit2005field}. The theory is called {\em renormalizable} if this number is finite. The same applies to higher-order contributions to the noise \cite{Hinrichsen00}. All terms under the RG flow vanish are called {\em irrelevant} and can, therefore, be safely neglected to study the system's behavior around the critical points. {\em Marginal} terms introduce logarithmic corrections to the theory obtained by neglecting their contribution. For instance, for the Ising and CP cases, the corresponding Langevin equations, excluding terms that are irrelevant at all spatial dimensions, can be written as
\begin{equation}
\begin{cases}
    \dot{\phi} = \mu^2\phi + \lambda\phi^3 + D\nabla^2\phi + \xi(\mathbf{x},t) & \text{(Ising)} \\
    \dot{\phi} = a\phi - b\phi^2 + D\nabla^2\phi + \sqrt{\phi} \xi(\mathbf{x},t) & \text{(CP)}
\end{cases}
\label{MFEq}
\end{equation}
which define two of the most studied universality classes --with characteristic critical exponents-- in statistical physics that stand as fundamental cornerstones upon which nearly any model in this field is built.

What is, then, the core structure of the RG approach in statistical physics? 
It consists of a theoretical framework to determine the scaling behavior of the thermodynamic features of a physical system when subjected to coarse-graining and rescaling of the lengths in the neighborhood of a second-order phase transition where the only physically relevant scale is the correlation length $\xi$ much larger than the microscopic one. Indeed, the coarse-graining step integrates out the small-scale features of the system from the lower cut-off $\epsilon$ up to an arbitrary limit $\epsilon'>\epsilon$ (smaller than $\xi$). In contrast, the rescaling step changes the length unit to reestablish the original value of the lower cut-off, i.e., $\epsilon'$ is contracted to $\epsilon$. In this way, the original correlation length of the system is also reduced by the scaling factor $\epsilon/\epsilon'$. The only case in which the correlation length is unchanged is when it diverges, i.e., the system is scale invariant. This situation defines a fixed point of the RG transformation coinciding with a second-order phase transition. From the velocity in which the interaction parameters of the field theory change under these two RG steps, one can derive the scaling behavior of the thermodynamic features of the system around the {\em critical} fixed point. 

In principle, the RG procedure can be performed in real space (Kadanoff's RG) or in the conjugate Fourier space (Wilson's RG), with this second solution more suitable for quantitative calculations. 

\begin{figure}
    \centering
    \includegraphics[width=0.9\columnwidth]{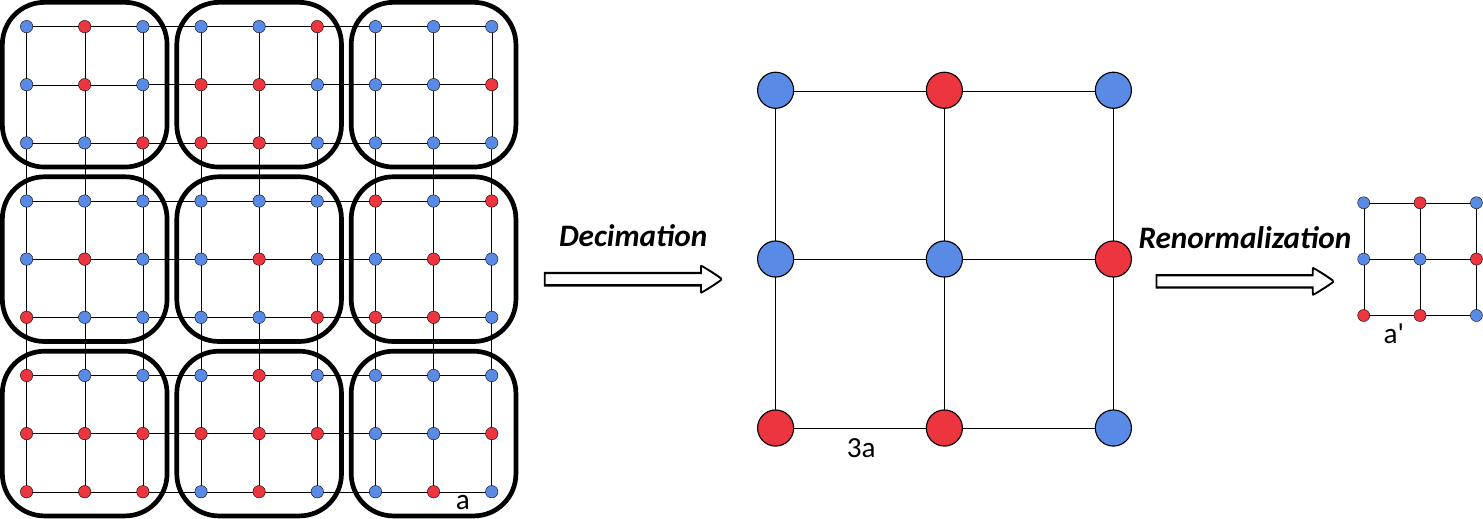}
    \caption{Sketch the decimation process employing Kadanoff blocks on a square lattice of side $a$, where blue (red) points represent up (down) spins or active (inactive) sites,  respectively. The lattice is divided into several blocks with $b^2$ sites, which are now coarse-grained, replacing it with a single block following some rule $\mathcal{R}_g(\sigma)$, finally reducing all the system scales by a factor $b$. This scheme produces a reduced version of the original system.}
    \label{Kdnff}
\end{figure}

Let us start with the real space RG. It can be performed either directly on the lattice model (e.g. Ising model) or in the field theory representation (e.g., Eq.~\eqref{Ising}. Kadanoff originally proposed the real space coarse graining idea in 1966 \cite{migdal1976phase, kadanoff1966scaling} for the lattice Ising model and follows the subsequent three-step procedure (also schematized in Figure \ref{Kdnff}). This defines the RG process \emph{\`a la Kadanoff}:
\begin{enumerate}
 \item Group the lattice points into groups of $b^d$ sites in identical $d-$dimensional cubic cells. This is possible thanks to the homogeneity and translational invariance of the embedding space and the Hamiltonian.
 \item Replace each block of micro-spins $s_i$ with a single macro-spin $\sigma_j$, adopting some physically meaningful rule to associate a value of $\sigma_j$ (e.g., $\pm 1$) to each configuration of the micro spins in the cell (e.g., a majority rule). Write the Hamiltonian for the macro spins appropriately, transforming the coupling constants and eventually considering the appearance of higher-order interactions. 
 \item Rescale all lengths by a factor $b$ to return to the original lattice spacing and repeat all the steps.
\end{enumerate}
The transformation equations of the interaction constants of the model define the RG flow whose fixed point represents the scale-invariant condition related to the critical point.

While the above ``real space" version of RG is conceptually very clear, it is rarely accurate enough for quantitative calculations, which are more conveniently performed within the $k-$space Wilson's formulation of the RG \cite{wilson1974renormalization}. In order to briefly illustrate it, let us start from the partition function of the field theory defined by the following functional integral,
\begin{equation}
    Z=\int\mathcal{D}\phi(\mathbf{x},t)e^{-\mathcal{F}[\phi(\mathbf{x},t)]}
    \label{PartFunc}
\end{equation}
where $\mathcal{F}[\phi(\mathbf{x},t)]$ is the Hamiltonian or the Lagrangian function of the field theory. For instance, for the Ising model, it is given by Eq.~\eqref{Ising} (and $\phi(\mathbf{x},t)$ does not depend on time). The partition function $Z$ determines all the thermodynamic properties of the system.
Let us call $\mathcal{F}_G[\phi(\mathbf{x},t)]$ the quadratic part of $\mathcal{F}[\phi(\mathbf{x},t)]$, giving rise to the Gaussian approximation of the theory. On a very general ground (see \cite{Hinrichsen00, Mam1997} and references therein), this is substantially determined by the Laplacian $\nabla^2$ operator of the embedding Euclidean space for the equilibrium models and by the heat equation operator $\partial_t-D\nabla^2$ for out-of-equilibrium models (e.g., contact process) which satisfy the simple requirement of locality of interactions and spatio-temporal homogeneity\footnote{Note that $\nabla^2$ is the simplest regular differential operator satisfying locality, symmetry by reflection and that does not contain an intrinsic scale length.}. 
Before applying the RG, it is convenient to use the Lagrangian on the basis that its exactly solvable Gaussian part takes a diagonal form. Due to the homogeneity of the Euclidean embedding space and the statistical translational invariant of the model, this is the ordinary Fourier $k-$space. Indeed, the plane waves $\exp[i\mathbf{k}\cdot\mathbf{x}]$ are the eigenfunctions of the translation operator and, consequently, of the Laplacian $\nabla^2$ with eigenvalues $-k^2$. The non-Gaussian part of the Lagrangian, in which different $k$ modes of the field $\phi$ are coupled together, is then considered in the framework of perturbation theory around the Gaussian approximation. This aspect is of great importance for the extension we propose of the RG approach to dynamical processes defined on irregular networks.
At this point, the RG approach proceeds along the following steps:
\begin{enumerate}
    \item  Integrate out high momentum modes, $\frac{\Lambda}{\zeta}<k<\Lambda$ where $\Lambda$ is the upper cut-off of the momenta $k$ (i.e., $\Lambda \sim 1/\epsilon$ of the real space approach) and $\zeta>1$ is a scale parameter. This implies a corresponding change of the coupling constants in the new Lagrangian defined for $k<\frac{\Lambda}{\zeta}$
    \item Rescale the momenta $\mathbf{k^\prime}=\zeta\mathbf{k}$ to reestablish the original upper cut-off for the momenta.
    \item Rescale the fields to normalize the gradient term (i.e., the Laplacian term of the Gaussian approximation).
\end{enumerate}
This defines the coupling constants as a function of the scale parameter $\zeta$. The fixed points are defined as before the critical points, and the behavior of the coupling constants around a critical point defines relevant and irrelevant coupling constants and the thermodynamic behavior of the system around the critical point.

Through the work of  Kadanoff in real space and Wilson in $k$-space, the Renormalization Group (RG) presents a sophisticated and accurate framework for studying criticality. This theory allows us to connect diverse spatiotemporal scales, comprehend the fundamental concept of scale invariance through the scaling hypothesis, and, finally, provide precise computing tools for the critical exponents in various spatial dimensions, utilizing the epsilon expansion as a perturbative approach (as outlined in the celebrated Wilson's paper "Critical Exponents in 3.99 Dimensions" \cite{Wilson1972}). Hence, one of the striking implications of the RG is the classification of many different dynamical models and physical systems --at criticality-- in a relatively small set of universality classes.

\subsection{Disorder and fluctuations: an RG perspective}
Many real-world processes, such as opinion dynamics, epidemic spreading, or economic shocks, can be modeled as dynamic stochastic processes embedded in heterogeneous interactions among the elements involved. Many spreading processes, such as infections, do not have a uniform transmission rate or a regular topology of contacts. They can vary in space and time and often occur in non-uniform environments. Neuronal interactions in brain networks lead to synchronized states that regulate circadian rhythms or epileptic seizures, while opinion spreading can lead to polarization in social networks, or financial contagion in an interbank network may result in a large cascade of defaults. Most models introduced to study these phenomena have been initially defined on regular lattices with interactions satisfying the minimal locality and homogeneity requirements. All these systems are characterized by an irregular pattern of interactions, often showing hierarchical structures, strong heterogeneity of contacts, and small-world features. Hence, it is crucial to study how this persistent topological disorder of the processes' embedding space can impact the physical system's critical properties. In short, the problem can be reduced to the following question: What is the interplay between the dynamic process parameters and the presence of irregular topological features of the embedding space in determining transitions, as in the contact processes, from noisy inactive states to collective ordered ones (e.g., the transition from an inactive infection to a global epidemic state)?

A complete formulation of this question is still an open problem and thus goes beyond the scope of this manuscript. Here, we want to highlight some known facts on the effect of \new{the} introduction of quenched disorder in lattice models. Since in these out-of-equilibrium lattice models, ‘time’ and ‘space’ act differently on the process, it is customary to distinguish between spatial and temporal properties, denoting them respectively by the indices $\bot$ and $\Vert$. Indeed, non-equilibrium phase transitions are usually characterized by two independent correlation lengths: a spatial length scale $\xi_\bot$ and a temporal length scale $\xi_\Vert$ \cite{Hinrichsen00}.

What is known about \new{the} introduction of disorder in such lattice models is the following: in general, quenched disorder leads to irrelevant perturbations to the RG fixed point, as long as the system can be studied in the annealed approximation \cite{Hinrichsen00}. However, it has been demonstrated that the only introduction of temporal quenched disorder is generally a relevant perturbation for the directed percolation universality class in all dimensions. Consequently, the critical behavior and the associated critical exponents change entirely and vary continuously depending on the disorder strength. For instance, the temporal disorder has been shown to be relevant at the Ising transition only at and above three dimensions and to induce the so-called Temporal Griffiths phases (TGPs) characterized by generic power-law scaling of some magnitudes and generic divergences of the susceptibility \cite{TGP_MAM}. 

Similarly, the spatially quenched disorder is a marginal perturbation that may, therefore, seriously affect the critical behavior of the system at criticality depending on its specific spatial dimension (we refer to \cite{Votja2006} for a complete discussion of the issue). In particular, the quenched spatial disorder can lead to a singularity in the free energy, thus having dramatic consequences for the properties of continuous phase transitions and generating rare region (RR) effects characterized by generalized slow dynamics: the so-called Griffiths phases \cite{Griffiths1969}.

However, when dealing with dynamical models defined on strongly irregular networks, the topological disorder due to the embedding space can no longer be treated as a perturbation concerning the homogeneous lattice case. A new approach has to be formulated, as shown below.

\subsection{Zooming out heterogeneous networks}

In order to deal with the case of strong spatial disorder, as in the case of a complex heterogeneous networked embedding space, let us start with some considerations. As shown above, about the $k-$space RG approach, an important point to bear in mind is that Gaussian field theories can always be exactly solved, and in general, the related interactions are governed by the Laplace differential operator. Moreover, the Gaussian approximation works well above the upper critical dimension $d\ge d_c$. Therefore, an important starting point of the RG {\em à la} Wilson is to write the model on the basis that diagonalizes such Gaussian approximation, i.e., the basis of eigenstates of the Laplacian operator (in homogenous spaces, these are the plane waves).
The Gaussian model on random graphs has been studied in \cite{Cassi1, Cassi2}. In this case, the Hamiltonian, in analogy with the Gaussian approximation of Eq.~\eqref{Ising}, can be rewritten as 
\begin{equation}
    \mathcal{H}=\frac{1}{2}\underset{ij}{\sum}\phi_{i}\left(L_{ij}+m_{i}^{2}\delta_{ij}\right)\phi_{j}\,,
\end{equation}
where $\phi_i$ is a real field, and $m_i^2=\alpha_i m^2$, with $1/K<\alpha_i<K$ for some positive $K$ representing the square masses. Note that, now, $\hat L=\hat D - \hat A$ is the analog of the continuous Laplace operator $(-\nabla^2\phi)$ (by historical reason in graph theory, the Laplacian operator is the network generalization of $-\nabla^2$ instead of $\nabla^2$), defined on a graph or a discrete grid, being $\hat A$ the adjacency matrix of the network and $\hat D$ the diagonal degree matrix of the network.

If we want to extend the RG approach to models with identical nearest neighbor interactions but defined on heterogeneous graphs, the fundamental step is a physically meaningful formulation of a correct coarse-graining procedure, either in real
space, through the derivation of mesoscopic collective variables (i.e., block variables) from a local resummation of the microscopic ones inside `equivalent' mesoscopic cells tiling the whole space, or in the conjugate `Fourier' space by rewriting the model in an suitable `Fourier'
basis and explicitly integrating all the `modes' with `wavelength' smaller than an arbitrary mesoscopic determined by a scale factor $s$. In both cases, the new lower cut-off of the model is given, respectively, by $s$. 
As aforementioned, given a homogeneous model (i.e., with interaction constants identical for all interacting neighboring pairs,
as the Ising model with nearest-neighbor interactions, for instance), the definition of coarse-graining in identical mesoscopic cells or integrating small wavelength modes all over the space is immediate both in homogeneous spaces (e.g., $\mathbb{R}^{d}$) or homogeneous lattices or trees due to the continuous or discrete translational invariance of the space itself. This fundamental property is completely lost in heterogeneous networks. Consequently, both the definition of real space block variables and the resummation procedure of small wavelengths of the model in Fourier space may appear completely arbitrary or even meaningless.

Our Laplacian scheme for connected undirected graphs tackles this point as a natural extension of the coarse-graining in homogeneous spaces. First, let us notice that the Laplacian operator $\nabla^{2}$ in such spaces can be strictly related to the progressive coarse-graining procedure in the RG: it is an operator with the spectrum of eigenvalues given by eigenvalues $-k^{2},\ $with \emph{k} Fourier wave-vector, i.e., the inverse of the wavelength, which has no characteristic scale. The corresponding eigenvectors are exactly given by the Fourier basis of plane waves $e^{\text{ikx}}$, which are also eigenfunctions of the translation group. In other words, the Laplacian operator is a telescopic ``scanner'' of the coarse-graining scales. Second, the diffusion equation $\partial_{t}\rho=-\hat{L}\rho$ with $\hat{L}=-\nabla^{2}$ couples the spatial scales ``scanned'' by $\nabla^{2}$ to corresponding diffusion times, which are proportional to the square of the wavelengths or, inversely, at each value of the time. The diffusion, therefore, reaches from each point an identical length of the order of the square root of the time itself, defining in this way identical coarse-graining cells around each point \emph{à la} Kadanoff. In other words, the coarse-graining \emph{à la} Wilson in a homogeneous space can be performed by integrating out all wavelengths smaller than the square root of an arbitrary diffusion time and then increasing this time.

The symmetric graph Laplacian operator $\hat{L}=\hat{D}-\hat{A}$
is the natural graph discrete representation of the continuous operator
$-\nabla^{2}$. It exactly describes the same diffusion dynamics $\dot{\rho}=-\hat{L}\rho$ on a heterogeneous connected network with the difference that for a given time, such dynamics cover different structures at different locations on the network due to the connectivity heterogeneity of the space. However, these structures share the fundamental property of being covered by diffusion simultaneously. In this strict sense, our formulation of Laplacian RG for graphs can be seen as the natural extension to heterogeneous networks of the usual RG approach in statistical physics and statistical field theory. Being impossible to define identical coarse-graining cells or an integration scheme over small wavelengths due to the lack of spatial translational invariance, i.e., due to the topological inhomogeneity of the space, we can, however, adopt the Laplacian operator and the diffusion equation as a tool to define the coarse-graining procedure both on homogeneous spaces and on inhomogeneous graphs.

It is noteworthy that LRG is optimal for intrinsic scale detection and scale transformations in dynamical processes defined on heterogeneous networks also for the following reason: most of these processes (e.g., Ising, contact, epidemic, Kuramoto, and voter models) when defined on regular lattices, where interactions between nodes satisfy minimal requirements of locality and homogeneity, give rise to statistical field theories in which the Laplacian operator of the embedding space defines the Gaussian approximation \cite{Hinrichsen00, Dornic2005}. The RG analysis of these models, therefore, develops along the following steps: (i) exactly solving the Laplacian/Gaussian approximation, writing the theory on the basis on which the Gaussian theory is diagonal, i.e., based on eigenvectors of the operator $\hat L$; (ii) taking into account higher power terms in the framework of perturbation theory; (iii) integrating out the contribution of larger and larger scale eigenvectors of the Gaussian kernel, appropriately rescaling the length scales at each step. These are the steps to develop a full LRG theory to define an RG approach to irregular networked spaces.

\section{Results: The Laplacian Renormalization Group}
The first step is, thus, to develop a successful coarse-graining method for heterogeneous systems relying on the definition of a canonical formulation of information diffusion in heterogeneous environments. This can be based on the communication of information in complex networks whose evolution is governed by the Laplacian matrix  \cite{newman2010networks,masuda2017random}, defined for general weighted but undirected networks as 
\begin{equation}
L_{ij}=[ (\delta_{ij}\sum_k A_{ik})-A_{ij}],
\label{Laplacian}
\end{equation}
where $A_{ij}$ are the elements of the adjacency matrix $A$, and $\delta_{ij}$ is the Kronecker delta function. 

Note that, using Eq.\eqref{Laplacian}, we can now analyze how any scalar field evolves with time from a given initial specific state  $\mathbf{\phi}(0)$. This can be written as  $\mathbf{\phi}(\tau)=e^{-\tau \hat L}\mathbf{\phi}(0)$, where the temporal evolution of $\phi$ will depend on the \emph{network propagator},  
\begin{equation}
\hat K=e^{-\tau \hat L},
\end{equation}
representing the discrete counterpart of the path-integral formulation of general diffusion processes \cite{feynman2010quantum,zinnjustin2007phase}, where now each matrix element $\hat K_{ij}$ describes the sum of diffusion trajectories along all possible paths connecting nodes i and j at time $\tau$ \cite{moretti2019network,Cassi2}.
In terms of the network propagator, $\hat K$, it is possible to define the ensemble of accessible information diffusion states \cite{dedomenico2016spectral,ghavasieh2020statistical,InfoCore}, namely,
\begin{equation}
 \mathbf{\hat \rho(\tau)}= \frac{\hat K}{\mathrm{Tr}(\hat K)}=\frac{e^{-\tau \hat L}}{\mathrm{Tr}(e^{-\tau \hat L})}\,.
 \label{EvolMat}
\end{equation}
where $\hat \rho(\tau)$ is tantamount to the canonical density operator in statistical physics (or to the functional over fields configurations) \cite{binney1992theory,pathria2011statistical,greiner2012thermodynamics}. We assume connected networks fulfill the ergodic hypothesis. 

At this point, it is natural to define the canonical system entropy \cite{dedomenico2016spectral,InfoCore}, $S[\hat \rho(\tau)] = -\operatorname{Tr}[\hat{\rho}(\tau) \log \hat{\rho}(\tau)]$. Indeed, the temporal derivative of the entropy represents the heat capacity of a network \cite{InfoCore}, 
\begin{equation}
    C(\tau)=-\frac{dS}{d(\log \tau)},
    \label{SHeat}
\end{equation}
being the counterpart of the specific heat in classical statistical mechanics and linked to the correlation length in the system. In particular, $C$ is a detector of structural transition points corresponding to the intrinsic characteristic diffusion scales of the network \cite{InfoCore,villegas2023laplacian}. Indeed, a pronounced peak of C defines $\tau=\tau^*$ and reveals the starting point of a strong deceleration of the information diffusion, separating regions sharing a rather homogeneous distribution of information from the rest of the network. If more well-separated diffusion timescales exist, $C(\tau)$ can show a multi-peak structure.

In particular, as $S\in[0,1]$, it can reflect the emergence of \emph{entropic transitions} (or information propagation transitions, i.e., diffusion) over the network \cite{InfoCore}. 
Indeed, by increasing the diffusion time $\tau$ from $0$ to $\infty$,  $S[\hat \rho(\tau)]$ decreases from $1$ (\emph{segregated} and heterogeneous phase -- the information diffuses from single nodes only to the local neighborhood) to $0$ (\emph{integrated} and homogeneous phase -- the information has spread all over the network).

\begin{figure}[hbtp]
    \centering
    \includegraphics[width=1.0\columnwidth]{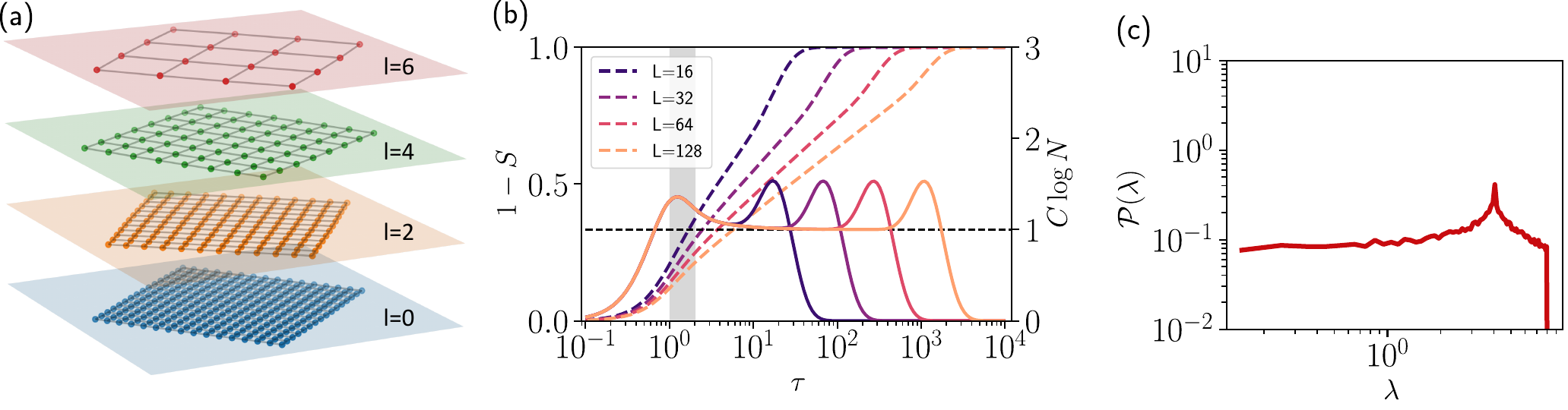}
    \caption{(a) LRG process for a 2D lattice of size $N=16^2$ employing $\tau=1.5$. Each layer represents the LRG step of order $l$. (b) Entropy parameter for the 2D lattices considering different lattice sizes ($N=L^2$, see legend). The first peak allows us to identify the characteristic network scale to perform the coarse-graining process. (c) Probability distribution of Laplacian eigenvalues for a 2D lattice of size $N=128^2$. The spectrum presents a characteristic scale at $\lambda=4$ (the number of neighbors), while the flat region leads to the constant value $C=1$.}
    \label{2DLatt}
\end{figure}

For the sake of clarity, we present here the simplest trivial case: a regular two-dimensional lattice. As shown in Figure \ref{2DLatt}(b), the smallest $\tau^*$ peak reflects the characteristic resolution scale of the system: it is related to the cut-off $\Lambda$, opening the door to formulate the LRG framework. Figure \ref{2DLatt}(a) shows several LRG steps for this trivial case (see below)). Note that fine-grained transformations are consistent over all the greyish areas of Figure \ref{2DLatt}(b), after which, in small networks, Kadanoff supernodes may become too large, and finite-size effects can alter the renormalization process.

\subsection{Real-space formulation}\label{RealSpace}

The canonical formulation of a network presented above permits us to manage the characteristic system scales. We want to emphasize that the original real-space dynamical RG scheme requires
the consideration of two fundamental scales: the lattice space $(a)$ and the correlation length of the system $(\xi_\bot)$. In concomitance with the original formulation, the peaks in the specific heat of the information diffusion flow allow us to identify the characteristic scales of the system: they are the counterpart to the correlation length or the lattice spacing when the process is carried out over blocks of spins or active sites in percolation \cite{christensen2005complexity}.

\begin{figure}[hbtp] 
\begin{center}
 \includegraphics[width=1.0\columnwidth]{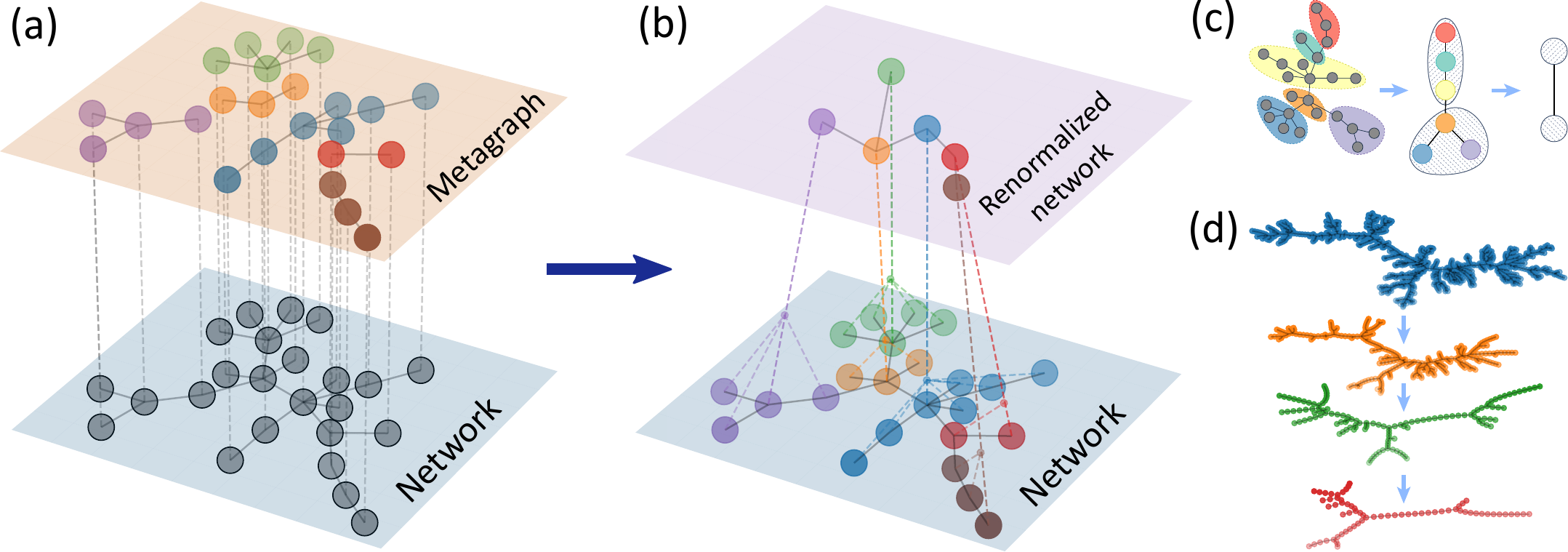}
 \caption{Sketch of the Kadanoff supernodes procedure. (a) The lower layer shows the case of a BA network ($N=24$, $m=1$), and the upper layer, $\zeta$ for $\tau=1.96$. Different colors identify the Kadanoff supernodes. (b) Each block becomes a single node incident to any edge to the original ones. (c) Scheme for a three-step LRG process. (d) LRG transformations for a random tree.}\label{MetaG}
 \end{center}
\end{figure}

One of our main points is to define how to extract the network 'building blocks', i.e., to generate what we call a \emph{metagraph} at each time $\tau$, to link the different network mesoscales. 
Note that, when $\tau=0$, $\rho$ is the diagonal matrix $\rho_{ij}(0) = \delta_{ij} /N$. Hence, $\hat \rho(\tau)$ will be subject to the properties of the network Laplacian, ruling the current information flow between nodes and reflecting the renormalization group flow. So far, we need to consider a rule (in a similar way to the 'majority rule') to scrutinize the network substructures at all resolution scales (i.e., $\tau$). 
For the sake of simplicity, we choose the following one: two nodes reciprocally process information when they reach a greater than or equal value than the information contained on one of the two nodes  \cite{InfoCore}, thereby introducing
\begin{equation}
\rho^\prime_{ij}=\frac{\rho_{ij}}{\min(\rho_{ii},\rho_{jj})}.
\end{equation}
\new{This is expected to give the set of minimal disjoint blocks in the system. Also, to simplify the network reduction process yet and,} depending on the particular $\rho_{ij}$ matrix element at time $\tau$, it is possible to define a metagraph, $\zeta_{ij}=\Theta(\rho'_{ij}-1)$, where $\Theta$ stands for the Heaviside step function. As expected, for $\tau \to \infty$, $\rho$ converges to $\rho_{ij}=1/N$, and $\zeta$ becomes the all-ones matrix. For a given scale, the metagraph $\zeta$ is thus the \emph{binarized counterpart} of the \new{disjoint set of nodes as given by the }canonical density operator \new{at some specific time $\tau$}. Note that, after examining all continuous paths traveling along the network \cite{zinnjustin2007phase} and starting from node $i$ at time $\tau=0$, our particular choice selects the most probable paths from  Eq.(\ref{EvolMat}), giving information about the prominent information flow paths of the network in the interval $0<t<\tau$. In statistical mechanics, we are considering the analogy to the Wiener integral and building the RG diffusion flow of the network structure \cite{wilson1974renormalization, zinnjustin2007phase}. The last step is to recursively group the network nodes into subsequent supernodes, i.e., how to perform decimation. Fig.\ref{MetaG} shows a snapshot of the procedure.

In full analogy with the Kadanoff picture, it is possible to consider nodes --under the accurate selection of particular blocking scales of the network-- within regions up to a critical mesoscale, which behaves like a single supernode \cite{kadanoff1966scaling, christensen2005complexity}. Analogously to the real-space RG, there is no unique way to generate new groups of supernodes or coarse-graining. For example, in a 2D squared lattice, there is no general rule about the size $\ell$ of the squared Kadanoff blocks to perform decimation and rescaling. If the system is scale-invariant, we expect it to be unaffected by RG transformations. In this perspective, using the specific heat, $C$, we propose an RG rule over scales $\tau\sim\tau^*$, where $\tau^*$ stands for the $C$ peak at short times, realizing the small network scales and dividing the network into the smallest possible supernodes. 

Therefore, as a general rule, we propose the following scheme \ 'a la Kadanoff \cite{villegas2023laplacian}: 
\begin{itemize}
    \item Build a network meta graph for $\tau\sim\tau^*$, i.e., a set of heterogeneous disjoint blocks of $n_i$ nodes.
    \item Replace each block of connected nodes with a single supernode.
    \item Consider supernodes as a single node incident to any edge to the original $n_i$ nodes.
    \item Realize the scaling {\em i.e.} write the Laplacian matrix for the new graph.
\end{itemize}

\begin{figure}[hbtp]
\centerline{
\includegraphics[width=1.0\columnwidth]{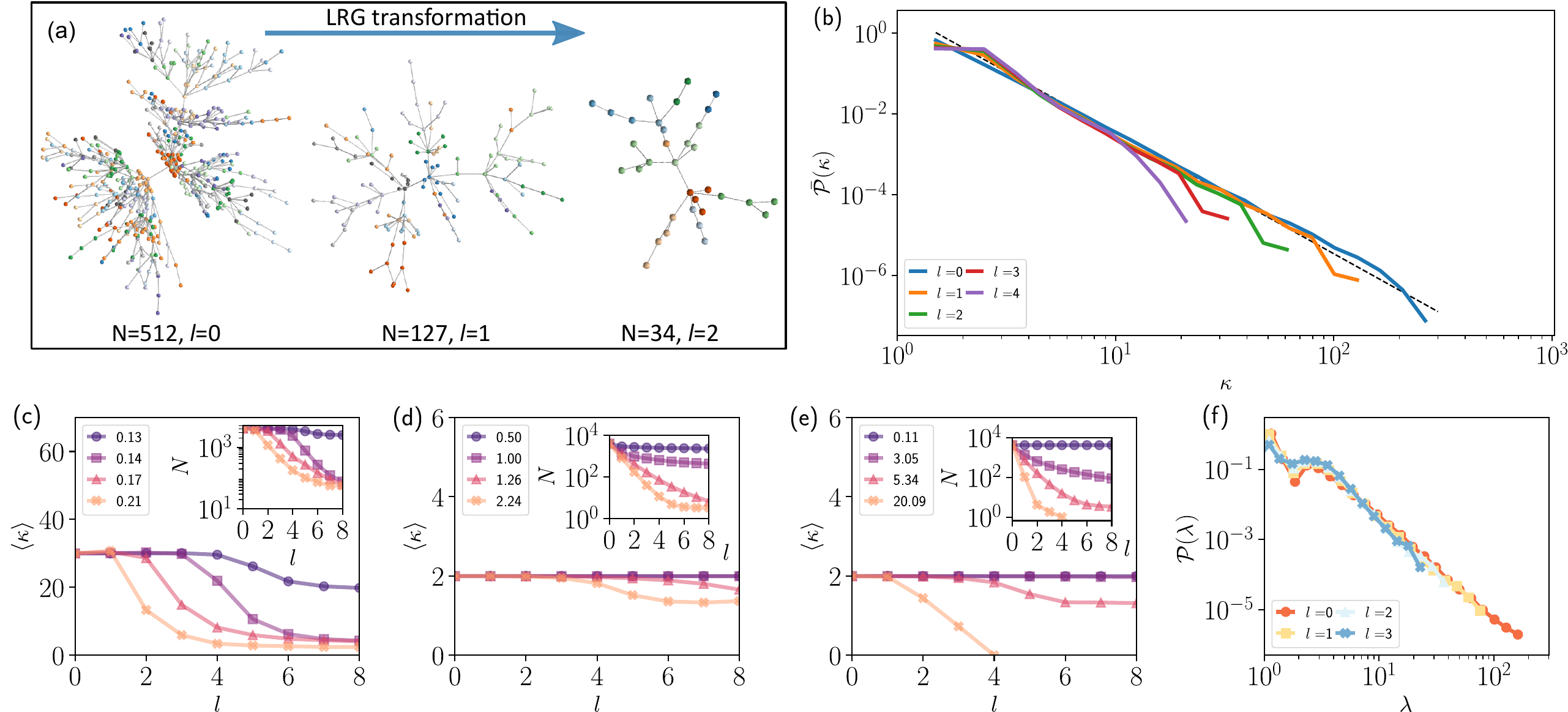}
}
\caption{\textbf{LRG coarse-graining. (a)} LRG transformation for a particular selection of a BA network ($N=512$, $m=1$). Kadanoff supernodes are plotted in a different color for every scale. \textbf{(b)} Degree distribution for a BA network (solid lines), with a characteristic exponent $\gamma=3$ (dashed line) at different RG steps with $\tau=1.26$ (see legend). \textbf{(c-f)} Mean connectivity flow under subsequent LRG transformations for different $\tau$ values (see legend): \textbf{(c)} an Erd\H{o}s-R\'enyi network of $\langle \kappa \rangle_0 =30$,  \textbf{(d)} a BA scale-free network with $m=1$ and, \textbf{(e)} random tree. \textbf{(g)} Spectral probability distribution, $\mathcal{P} (\lambda)$, of the downscaled Laplacian replicas for different LRG steps in a BA network (see legend). All curves have been averaged over $10^2$ network realizations with $N_0=4096$.}\label{KBlocks}
\end{figure}

Figure \ref{KBlocks} shows the application of multiple steps $l$ of the LRG over different networks. Erd\" {o}s-R\'enyi networks exhibit only a characteristic resolution scale. In other words, such a network cannot maintain its intrinsic properties under any coarse-graining transformation and, consequently, present a continuous flux through the collapse of the network, i.e., to a single-node state reflecting the existence of a well-defined network scale. Hence, for any possible grouping of nodes --at every scale, $\tau$-- the mean connectivity of the network decreases after successive RG transformations. On the contrary, random trees and Barab\'{a}si-Albert networks maintain their intrinsic properties when performing network reduction, as shown in Figure \ref{KBlocks}(d) and \ref{KBlocks}(e).

We refer to the original work presenting the LRG procedure \cite{villegas2023laplacian} for further analysis of coarse-graining applied to different scale-free real networks, i.e. following bonafide finite-size scaling hypotheses \cite{serafino2021true}, and significant cases previously analyzed in other RG approaches \cite{garcia2018multiscale, rozenfeld2010small} for producing downscaled network replicas.

\subsection{$k$-space formulation}

All of the above explains how to have an operating procedure for the decimation and scaling part of RG in real space (inspired by the RG theory \emph{\'a la} Kadanoff \cite{kadanoff1966scaling}. In this section, we introduce a formulation of the Laplacian Renormalization Group (LRG), which can be connected with the field theory $k$-space RG approach pioneered by Wilson in statistical physics \cite{wilson1979problems}. This formulation leads to a Fourier-space version of Kadanoff's "supernodes" scheme at each LRG step, offering a deeper insight into this process. 

Without sacrificing generality, let us consider a scenario where we intend to renormalize the information diffusion on the graph up to a specified time $\tau^\prime$, effectively retaining only \emph{diffusion modes} at scales (times) larger than $\tau^\prime$. Note that the point where $C$ exhibits a peak for short times corresponds to the ultra-violet cutt-off $\Lambda$, which is related to the smallest possible characteristic scale of the system. In other words, the smallest possible network scale corresponds to the maximum eigenvalue $\lambda_{max}$: the choice of a diffusion time-scale $\tau\sim \frac{1}{\lambda_{max}}$ coincides with the finest possible resolution of the network structures. For the sake of clarity, we adopt the bra-ket formalism, where $\left<i|\lambda\right>$ denotes the projection of the Laplacian eigenvector $\ket{\lambda}$ onto the $i^{{th}}$ node of the graph. In this notation, we can represent $\ket{i}$ as a normalized $N$-dimensional column vector, with all components equal to zero except for the $i^{{th}}$ component, which is set to 1. In bra-ket notation, the Laplacian operator becomes $\sum_{\lambda}\lambda\ket{\lambda}\!\bra{\lambda}$. We then identify the $n<N$ eigenvalues $\lambda\ge =\lambda^\prime=1/\tau^\prime$ and their corresponding eigenvectors $\ket{\lambda}$. 

Hence, a single LRG step \'a la Wilson involves \cite{villegas2023laplacian}:
\begin{itemize}
\item Reducing the Laplacian operator to the contribution of the $N-n$ slower eigenvectors with $\lambda<\lambda^\prime$, denoted as $\hat L_0=\sum_{\lambda<\lambda^\prime}\lambda \ket{\lambda}\!\bra{\lambda}$.

\item Rescaling the time variable, transforming $\tau\to \tau'$ in such a way that $\tau^\prime$ becomes the unit interval in the rescaled time variable. This leads to a re-definition of the coarse-grained Laplacian as $\hat L^\prime=\tau^\prime\hat L_0$.
\end{itemize}

\begin{figure}[hbtp] 
\centering
 \includegraphics[width=1.0\columnwidth]{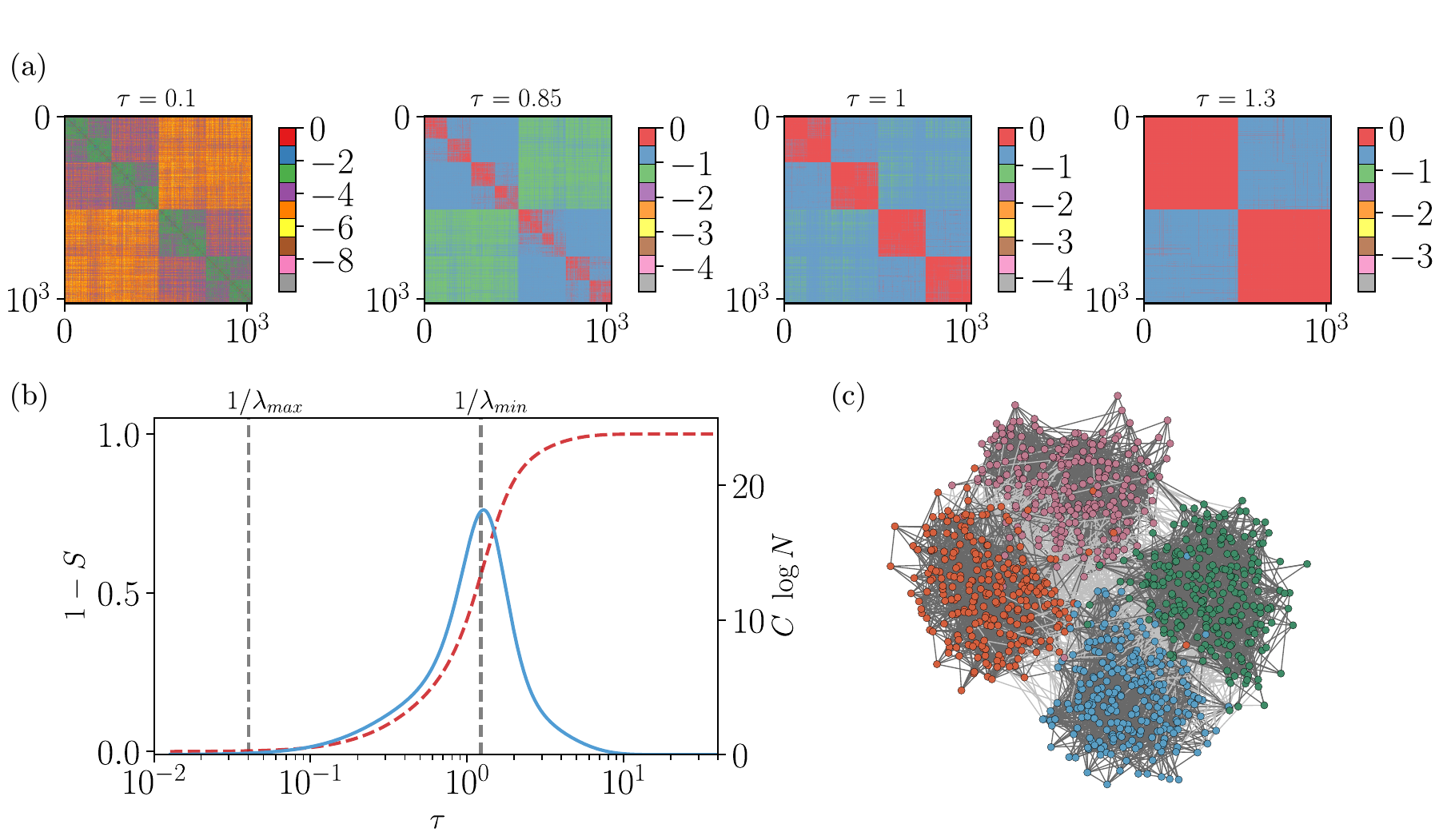}
 \caption{\textbf{Network defocusing.} (a) Temporal evolution of the logarithm of the density matrix (as written in Eq. \ref{CorrRho}). Once network eigenmodes are sequentially integrated, the different network mesoscales naturally emerge, making the different hierarchical modules used to build the network evident. (b) Entropy parameter (dashed lines, $(1-S)$)
and specific heat (solid lines, $C$) versus the temporal resolution parameter of the network, $\tau$. (c) The hierarchic modular network, with $N=1024$ nodes, that has been used for this specific example.}
\label{Density}
\end{figure}

Clearly, this real-space representation, as happens exactly in statistical physics, is just a mathematically approximated description of the $k$-space RG that preserves the physical meaning. We aim, however, to illustrate how different eigenvalues are canceled out, rendering distinguishable the different scales of the system. To do that, we consider here the following rescaled density matrix,

\begin{equation}
    \rho_0(\tau)=\log\rho(\tau)+\log(N)
    \label{CorrRho}
\end{equation}

where we have added the final term for the sake of comparison at different times. We want to remember that when $\tau=0$, $\rho$ is the diagonal matrix $\rho_{ij}(0) = \delta_{ij} /N$, and, for $\tau\rightarrow\infty$ $\rho$ is the full matrix with value $\nicefrac{1}{N}$. Figure \ref{Density}(a) shows the evolution of the density matrix for different times $\tau$, in the specific case of a random hierarchic modular network \cite{Zamora2016, Buendia2021}, showing the emergence of different network mesoscales when network eigenmodes are integrated out. \new{We want to emphasize that larger times also imply focusing on larger scales in the network. Therefore, small values of $\tau$ must be considered to create small Kadanoff supernodes.} We highlight how the diffusion time (see also Figure \ref{Density}(b))  effectively acts as a 'zoom lens,' evidencing the different network characteristic scales.

\subsection{Evolution of network metrics into the LRG flow}
A crucial consideration remains unresolved after defining a way to perform successive coarse-graining steps on heterogeneous structures: How do various network metrics scale for different topologies? Does our LRG scheme conserve the same characteristics of the networks as when well-known rules guide the network's growth?

We emphasize that the LRG framework conserves the probability distribution of the network Laplacian in $k-$space and coincides with RG transformations on lattices. This involves a sort of transformation that does not change the intrinsic properties of the network, and it is, thus, called an isospectral transformation \cite{Burioni1997}.

\begin{figure}[hbtp] 
\centering
 \includegraphics[width=1.0\columnwidth]{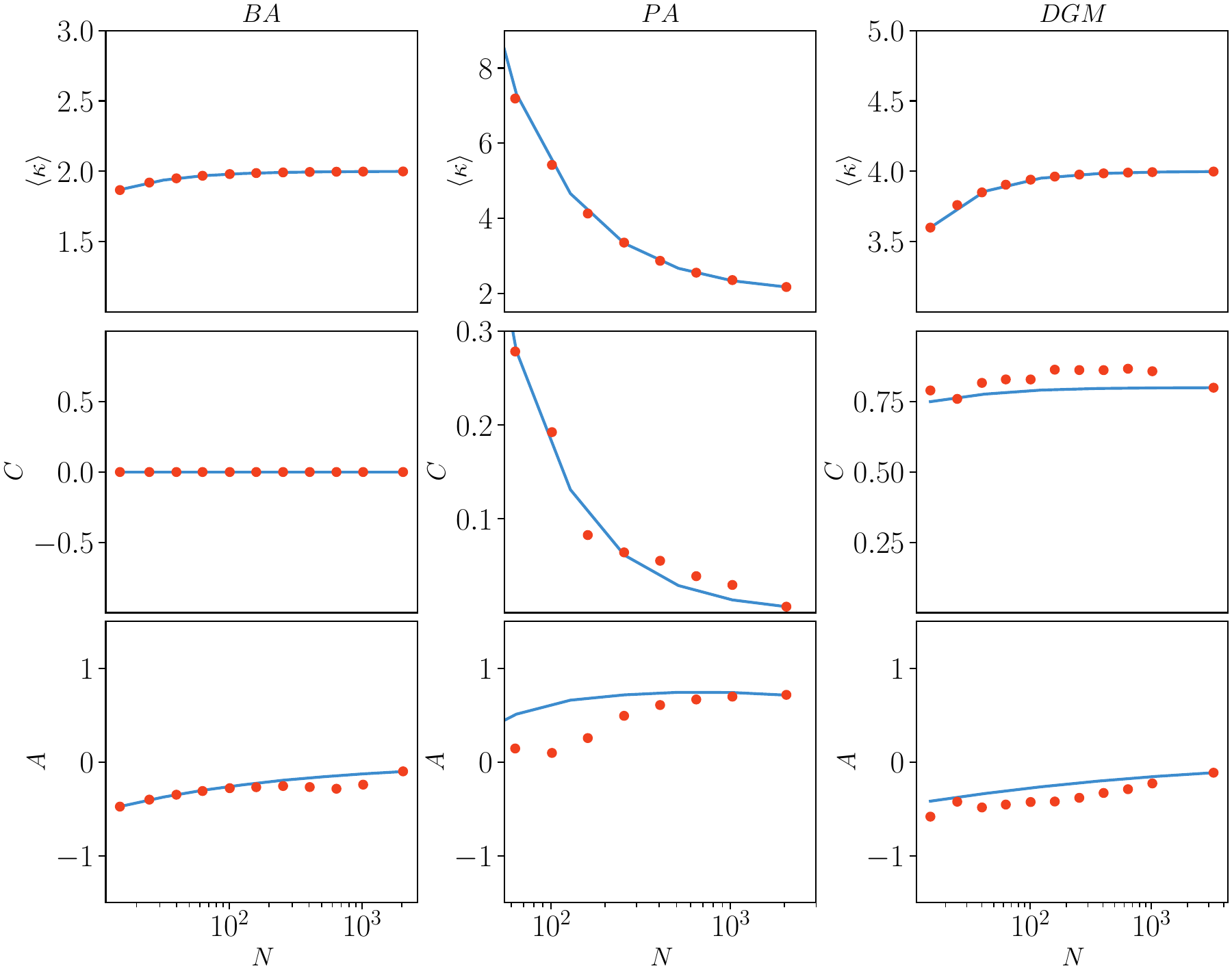}
 \caption{\textbf{Network observables.} Different observables versus system sizes for the growth process (blue lines) and LRG coarse-graining (red dots) for different synthetic networks: Barab\'asi-Albert (BA, left figures), Preferential Assortative (PA, central figures), and Dorogovtsev-Goltsev-Mendes (DGM, right figures). Note that in all three selected observables, the mean connectivity $\langle \kappa \rangle$, the average clustering coefficient $C$, and the mean assortativity of the network, $A$, the LRG fits with the expected theoretical line given by the network growth process. Parameters: The coarse-graining process has been done using $10^2$ averages on top of an initial network of $N=2048$ nodes for BA and PA and $N=3282$ for the DGM network. For the BA network, we have selected $m=1$, and for the PA network $M_0=20$ and $m=1$.}
\label{Metrics}
\end{figure}

Here, we have selected three particular network topologies: Barab\'asi-Albert networks (BA), Dorogovstev-Goltsev-Mendes networks (DGM), and preferential assortative (PA) networks to analyze how the clustering coefficient, the mean connectivity, and the network assortativity change when we perform a coarse-graining process of the selected networks.
In particular, we have designed the Preferential Assortative networks to grow in the following way: we start with a clique of $M_0$ nodes and attach each time step a new node with $m$ vertices as in the preferential attachment rule but considering the square of the probabilities to generate a sparser version of the network ($p_i=(\kappa_i/\sum_i\kappa_i)^2$). As observed in Fig. \ref{Metrics}, this simple choice generates networks with a high constant value of assortativity.

As shown in Figure \ref{Metrics}, we have selected different networks of fixed size, performing a coarse-graining process to reduce the network to very small sizes. Note that by selecting two of the networks that exhibit constant mean connectivity, a fixed value of the clustering coefficient, and zero assortativity on the infinite-size limit, we are able to follow different key points when performing heterogeneous coarse-grainings of different magnitude: the ability of the LRG to maintain the fixed connectivity, the constant clustering coefficient, and to predict finite-size effects as the convergence to the asymptotic value of the different network observables (see, for instance, the evolution of $A$ for BA and DGM networks). Instead, the choice of the PA network gives us important information in such cases where different variables evolve within the LRG flow in the growing process of the architecture. In any case, when performing the LRG process as explained in Section \ref{RealSpace}, we highlight the ability of the LRG to fit the growing process of well-behaved synthetic networks. That is, when the LRG performs an isospectral transformation, it maintains the flow path that the network has followed in the growth process. 

\begin{figure}[hbtp] 
\centering
 \includegraphics[width=1.0\columnwidth]{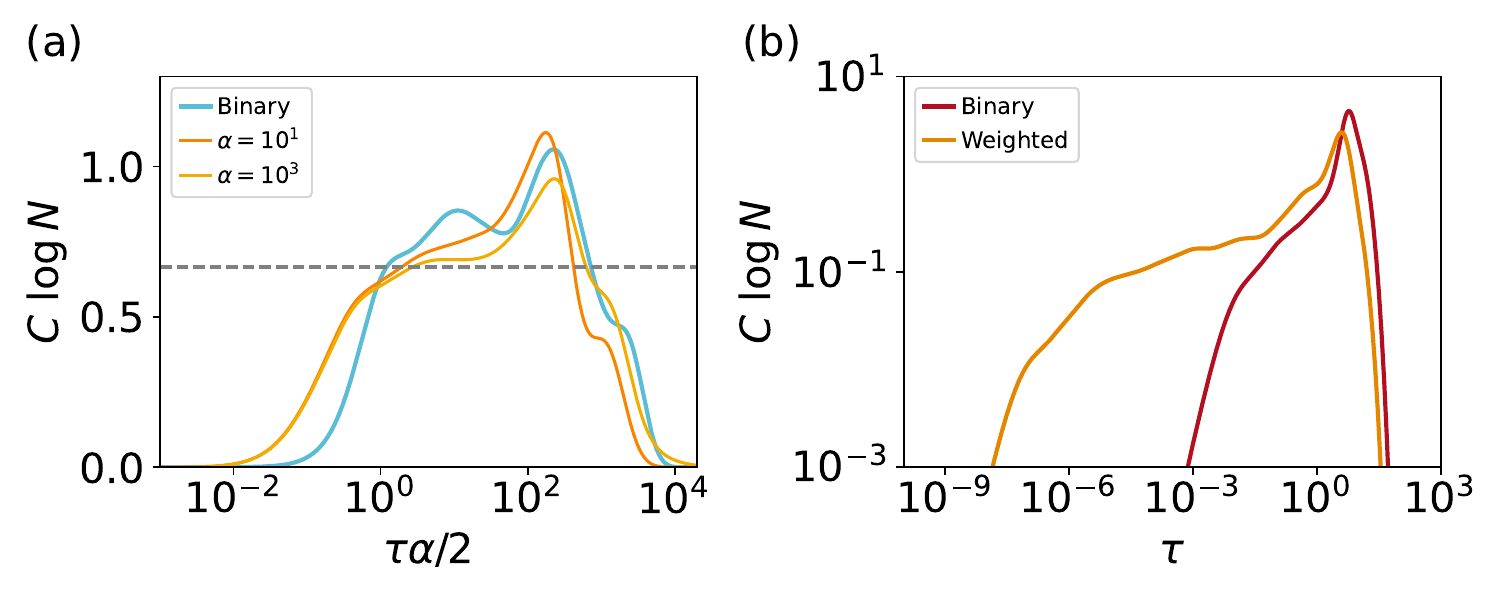}
 \caption{\textbf{Weighted networks.} Specific heat versus diffusion time for (a) the European road network between cities with different weights assigned randomly as a function of the node degree connecting the cities (see legend). The diffusion time has been rescaled by the mean weight for the sake of comparison. All curves have been averaged over $10^3$ independent realizations. (b) The USA airport network in its binarized and weighted version. Note that the microscopic heterogeneity, observable at short-length scales, is lost in the binarized version of the network. }
\label{Weighted}
\end{figure}

Once we have demonstrated that networks maintain their intrinsic properties due to the isospectral nature of the transformation \cite{Burioni1997}, we have also performed extensive simulations on a new application of the LRG method: the analysis of weighted networks. As can be deduced from the original definition, the Laplacian matrix of the network $\hat L = D-A$ can be used to perform coarse graining both in binarized or weighted networks. However, we emphasize that the intrinsic structure of weights on a complex network can change the nature of the observed heterogeneity of its binarized counterpart. Here, we present an application to weighted networks of particular interest: road networks. We have taken the European road network between cities \cite{subelj} and have computed the specific heat of two networks: the binarized one and a weighted version of the network considering the following rule for the different edges: we build a network ensemble where for each network realization the weight is randomly selected between $1$ and $\alpha$ times the maximum weight between the two cities representing the link. Therefore, we mimic the intensity of traffic as the weight interaction between the nodes $i$ and $j$. Note that the higher the city, the higher the possibility of having a large weight in the graph.  As shown in Figure \ref{Weighted} (a), the network road presents a multiscale structure for the binarized version, which ends in a network that looks like a random tree for high values of $\alpha$ with the selected weighted rule. Instead, Figure \ref{Weighted}(b) shows the weighted and binarized counterpart of the undirected airport network \cite{opsahl}, where each weight represents the total number of flights network between US airports (nodes) in 2010. As a direct consequence of this, we pinpoint that different weights will be canceled out in a different way in both cases, stressing the relevance of considering weights in real networks to scrutinize their intrinsic heterogeneity properly. 

\section{Discussion} \label{sec:IV}

Starting from the end of the 1960s and for more than 20 years, statistical physics has experienced significant scientific progress thanks to the development of the theory of critical phenomena and second-order phase transitions. It established the theoretical framework for many equilibrium and out-of-equilibrium statistical physical systems for the continuous transition when the system parameters are appropriately tuned from a disordered noisy phase to an ordered collective one through a critical point characterized by large fluctuations and long-range correlations \cite{kardar2007statistical}. The major step in the development of this theory has been the formulation of the Renormalization Group (RG) both in real and $k$-space \cite{kadanoff1966scaling,migdal1976phase,wilson1974renormalization}, which rigorously defines the operation of scale transformations and spatial rescaling to identify the relevant interactions of the system and quantify its behavior around the critical point where the only physically meaningful scale of the system is the large correlation length.

The RG has been developed under the hypothesis that the physical system's embedding space is homogeneous or translationally invariant (e.g., a regular lattice). Still, the theoretical problem of what happens when the homogeneity condition is violated has been studied for a long time by introducing a specific density of topological defects in homogeneous spaces/lattices. These studies can, therefore, be seen as the study of the effect of local “pathologies” of the homogeneous case. Only a marginal research activity was devoted at the same time to more complex cases, such as random graphs or trees \cite{Cassi1,Cassi2,Cassi3}.

The role of irregularities and topological heterogeneity in dynamical processes has become a central research field in the last twenty years with the advent of the complex networks research field as a natural paradigm for a vast class of real systems of great scientific interest \cite{Dorogovstev2008, Barzel2013}. They range from epidemic networks to biological and socio-economic systems and the human brain. For these systems, the complex network gives the skeleton of the interaction. It plays the role of the geometrically irregular space where the statistical dynamical processes are embedded (e.g., epidemic spreading or human brain dynamical activity). In all these systems, the highly inhomogeneous architecture of interactions cannot be treated as a simple topological perturbation of a homogeneous case but asks for a completely new theory that combines ab initio in a new theoretical approach to the complex geometrical structure of the space with the “physical” processes running on top of it. Due to the high topological complexity of the networked space, the concepts of criticality and phase transition themselves need to be reformulated in a more general sense that is suitable for this higher level of complexity.

Preliminary but essential steps in this direction have been recently developed by formulating a general approach to the problem \cite{InfoCore,villegas2023laplacian}: the LRG acts as a 'zoom lens' for all networks \cite{Klemm2023}. It grounds on two facts: (a) the Laplacian operator of a network describes the information diffusion dynamics on it so that it gives at each time the informationally equivalent neighborhood of each node of the network, proposing a natural extension to networks of the real space RG in homogeneous spaces \`a la Kadanoff; (b) the Laplacian RG can also be seen as a natural extension of the k-space RG \`a la Wilson in homogeneous spaces and practically all statistical dynamical models used to model physical processes on networks have a Gaussian approximation determined by the Laplacian operator \cite{villegas2023laplacian}. It solves, in a general way, the problem of coarse-graining the microscopic interacting elements into mesoscopic ones in an irregular network. Hence, the LRG has been shown to give immediate and profound implications in fostering groundbreaking new perspectives of network modularity \cite{Modularity}. Here, we have made further advances by demonstrating its capability to maintain the different network intrinsic properties (e.g., the clustering coefficient or the network assortativity) in synthetic networks where we know the growing rule of the system and under well-defined isospectral transformations. In particular, we have observed that the LRG can fit well with finite-size effects in all networks following the expected behavior when they grow or, in other words, allowing for a 'back and force' way to analyze networks in different fields. We emphasize that this possibility opens new avenues for extracting crucial information about different growing rules from real networks that are still unexplored.
Furthermore, we have shown for the first time the ability of the LRG to perform network reduction in weighted cases, leading to different reduced versions of the original network. However, expanding the LRG framework to various dynamic universality classes poses a significant challenge from a theoretical perspective. Despite the considerable progress made in this area, further research is still required to overcome this challenge \cite{NPEd2023}.

\section{Author contributions}
All authors contributed equally to the writing of this paper. All authors read and approved the final manuscript.

\section{Bibliography}
\bibliographystyle{unsrt}

\end{document}